# The Architectural Implications of Cloud Microservices

Yu Gan and Christina Delimitrou
Cornell University
{yg397,delimitrou}@cornell.edu

**Abstract**— Cloud services have recently undergone a shift from monolithic applications to microservices, with hundreds or thousands of loosely-coupled microservices comprising the end-to-end application. Microservices present both opportunities and challenges when optimizing for quality of service (QoS) and cloud utilization. In this paper we explore the implications cloud microservices have on system bottlenecks, and datacenter server design. We first present and characterize an end-to-end application built using tens of popular open-source microservices that implements a movie renting and streaming service, and is modular and extensible. We then use the end-to-end service to study the scalability and performance bottlenecks of microservices, and highlight implications they have on the design of datacenter hardware. Specifically, we revisit the long-standing debate of brawny versus wimpy cores in the context of microservices, we quantify the I-cache pressure they introduce, and measure the time spent in computation versus communication between microservices over RPCs. As more cloud applications switch to this new programming model, it is increasingly important to revisit the assumptions we have previously used to build and manage cloud systems.

**Index Terms**—Super (very large) computers, Distributed applications, Application studies resulting in better multiple-processor systems.

## 1 Introduction

Cloud computing services are governed by strict quality of service (QoS) constraints in terms of throughput and tail latency, as well as availability and reliability guarantees [8], [9], [10], [11]. In an effort to satisfy these, often conflicting requirements, cloud applications have gone through extensive redesigns [6], [7], [13]. This includes a recent shift from monolithic services that encompass the entire service's functionality in a single binary, to hundreds or thousands of small, loosely-coupled *microservices* [7], [21].

Microservices are appealing for several reasons. First, they simplify and accelerate deployment through modularity, as each microservice is responsible for a small subset of the entire application's functionality. Second, microservices can take advantage of language and programming framework heterogeneity, since they only require a common cross-application API, typically over remote procedure calls (RPC) or a RESTful API [1]. In contrast, *monoliths* make frequent updates cumbersome and error-prone, and limit the set of programming languages that can be used for development.

Third, microservices simplify correctness and performance debugging, as bugs can be isolated to specific components, unlike monoliths, where troubleshooting often involves the end-to-end service. Finally, microservices fit nicely the model of a container-based datacenter, with each microservice accommodated in a single container. An increasing number of cloud service providers, including Twitter, Netflix, AT&T, Amazon, and eBay have adopted this application model [7].

Despite their advantages, microservices change several assumptions we have long used to design and manage cloud systems. For example, they affect the computation to communication ratio, as communication dominates, and the amount of computation per microservice decreases. Similarly, microservices require revisiting whether big or small servers are preferable [5], [14], [15], quantifying the i-cache pressure from their code footprints [12], [18], and determining the sources of performance unpredictability across an end-to-end service's critical path. To answer these questions we need representative, end-to-end applications that are built with microservices.

In this paper we investigate the implications the microservices application model has on the design of cloud servers. We first present a new end-to-end application implementing a *movie renting, streaming, and reviewing system* comprised of tens of microservices. The service includes applications in different languages and programming models, such as node.js, Python, C/C++, Java/Javascript, Scala, and Go, and leverages open-source frameworks, such as nginx, memcached, MongoDB, Mahout, and Xapian [17]. Microservices communicate with each other over RPCs using the Apache Thrift framework [1].

We characterize the scalability of the end-to-end service on our local cluster of 2-socket, 40-core servers, and show which microservices contribute the most to end-to-end latency. We also quantify the time spent in kernel versus user mode, the ratio of communication to computation, and show that the shift to microservices affects the big versus little servers debate, putting even more pressure on single-thread performance. For the latter we use both power management techniques like RAPL on high-end servers, and low-power SoCs like Cavium's ThunderX. Finally, we quantify the i-cache pressure microservices induce, and discuss the potential for hardware acceleration as more cloud applications switch to this programming model.

## 2 Related Work

Cloud applications have attracted a lot of attention over the past decade, with several benchmark suites released both from academia and industry [12], [13], [17], [22], [23]. Cloudsuite for example, includes batch and interactive services, and has been used to study the architectural implications of cloud benchmarks [12]. Similarly, TailBench aggregates a set of interactive benchmarks, from web servers and databases, to speech recognition and machine translation systems, and proposes a new methodology to analyze their performance [17]. Sirius also focuses on intelligent personal assistant workloads, such as voice to text translation, and has been used to study the acceleration potential of interactive ML applications [13].

A limitation of current benchmark suites is that they exclusively focus on single-tier workloads, including configuring traditionally multi-tier applications like websearch as independent leaf nodes [17]. Unfortunately this deviates from the way these services are deployed in production cloud systems, and prevents studying the system problems that stem from the dependencies between application tiers.

## 3 The End-to-End Movie Streaming Service

We first describe the functionality of the end-to-end *Movie Streaming* service, and then characterize its scalability for different query types.

TABLE 1: Code composition of the *Movie Streaming* service.

| Service | Total New LoCs | Communication Protocol | LoCs for RPC/REST Handwritten | LoCs for RPC/REST Autogenerated | Unique Microservices | Per-language LoC breakdown (end-to-end service) |
|---|---|---|---|---|---|---|
| Movie Streaming | `10,263` | RPC | `8,858` | `46,616` | `33` | 30% C, 21% C++, 20% Java, 10% PHP, 8% Scala 5% node.js, 3% Python, 3% Javascript |

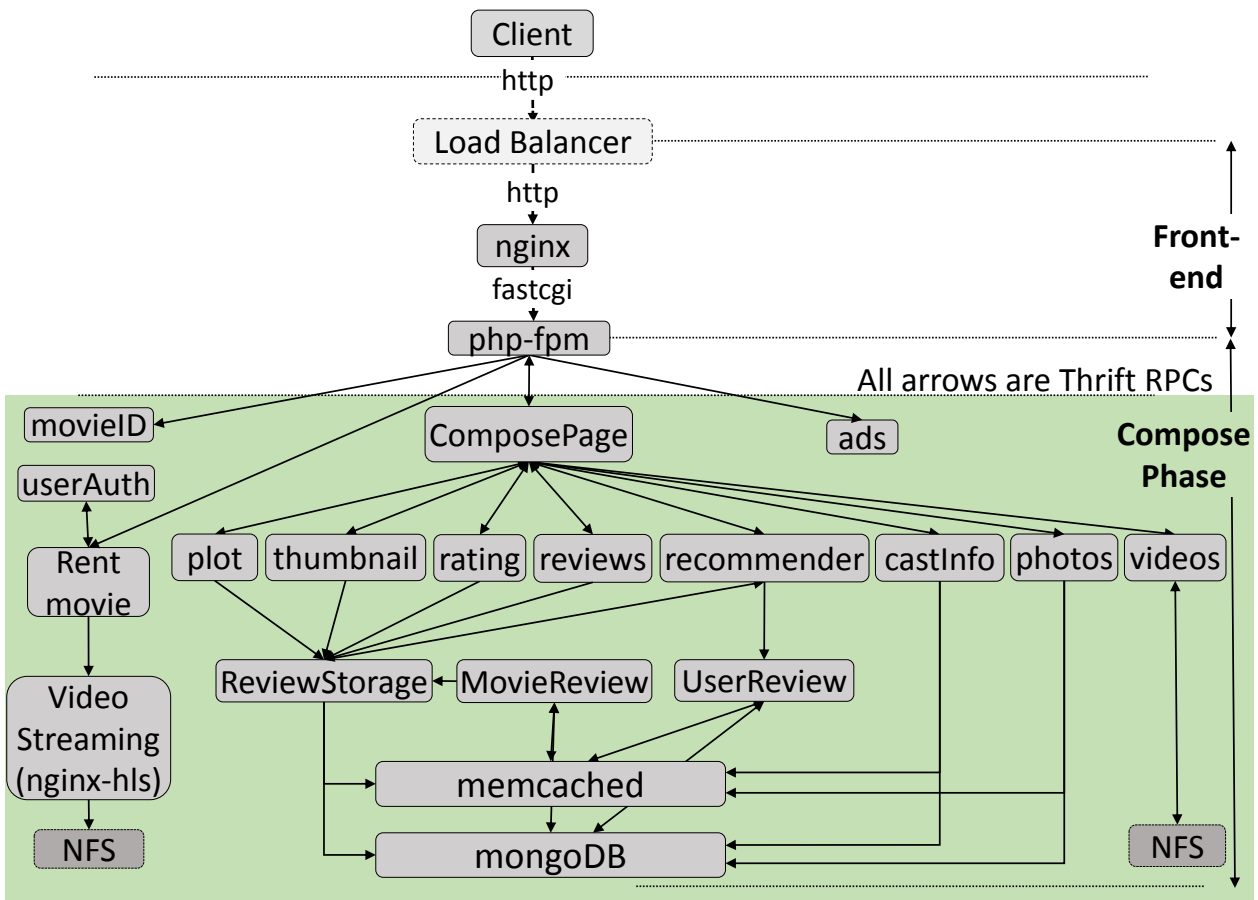

Fig. 1. Dependency graph for browsing & renting movies.

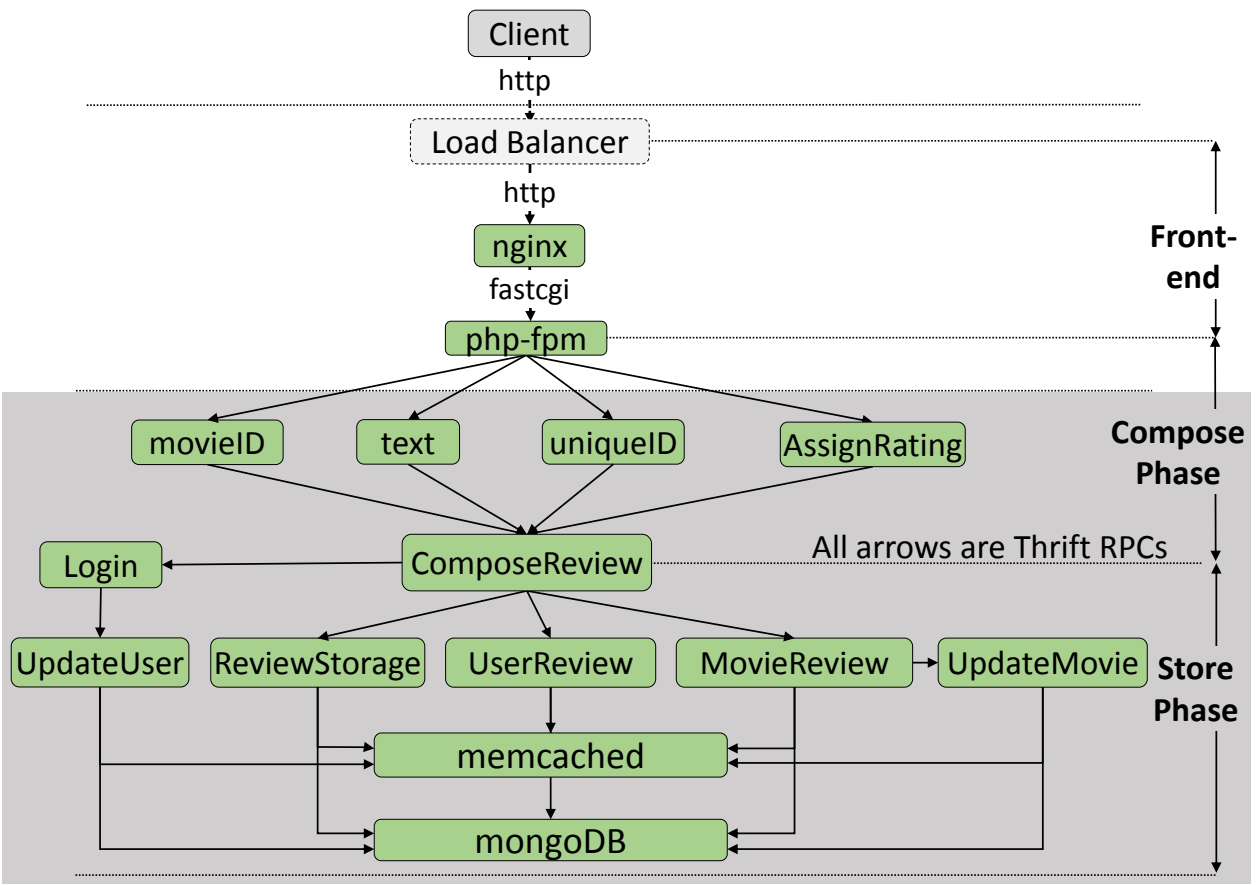

Fig. 2. Dependency graph for creating a new movie review.

## 3.1 Description

The end-to-end service is built using popular open-source microservices, including *nginx*, *memcached*, *MongoDB*, *Xapian*, and *node.js* to ensure representativeness. These microservices are then connected with each other using the Apache Thrift RPC framework [1] to provide the end-to-end service functionality, which includes displaying movie information, reviewing, renting and streaming movies, and receiving advertisement and movie recommendations. Table 1 shows the new lines of code (LoC) that were developed for the service, as well as the LoCs that correspond to the RPC interface; hand-coded, and auto-generated by Thrift. We also show a per-language breakdown of the end-to-end service (open-source microservices and RPC framework) which highlights the language heterogeneity across microservices. Unless otherwise noted, all microservices run in Docker containers to simplify setup and deployment.

**Display movie information:** Fig. 1 shows the microservices graph used to load and display movie information. A client request first hits a load balancer which distributes requests among the multiple `nginx` webservers. *nginx* then uses a `php-fpm` module to interface with the application's logic tiers. Once a user selects a `movieID`, `ComposePage` aggregates the output of eight Thrift microservices, each responsible for a different type of information; `Plot`, `Thumbnail`, `Rating`, `castInfo`, `Reviews`, `Photos`, `Videos`, and a `Movie Recommender` based on collaborative filtering [4]. The `memcached` and `MongoDB` instances hold cached and persistent copies of data on movie information, reviews, and user profiles, algorithmically sharded and replicated across machines. Connection load balancing is handled by the `php-fpm` module. The actual video files are stored in NFS, to avoid the latency and complexity of accessing chunked records from non-relational DBs. Once `ComposePage` aggregates the results, it propagates the output to the front-end webserver.

**Rent/Stream movie:** Thrift service `Rent Movie` uses an authorization module in *php* (`userAuth`) to verify that the user has sufficient funds to rent a movie, and if so, starts streaming the movie from disk via `nginx-hls`, a production nginx module for HTTP live streaming.

**Add movie reviews:** A user can also add content to the service, in the form of reviews (Fig. 2). A new review is assigned a $unique_{id}$ and is associated with a $movie_{id}$. The review can contain text and a numerical rating. The review is aggregated by `ComposeReview`, and propagated to the movie and user databases. `MovieReview` also updates the movie statistics with the new review and rating, via `UpdateMovie`.

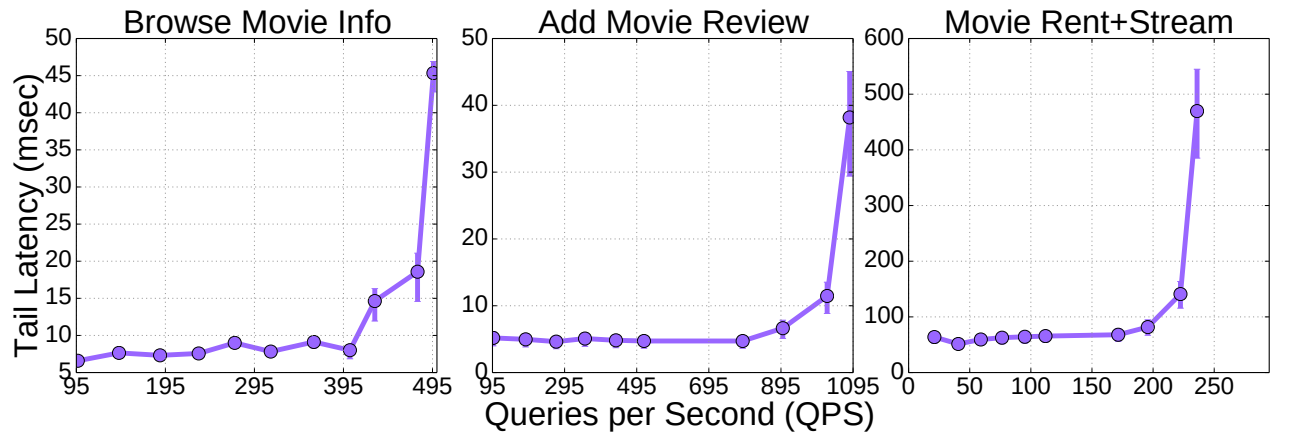

Fig. 3. Throughput-tail latency curves for different query types of the end-to-end Movie Streaming service.

Not pictured in the figures, the end-to-end service also includes movie and advertisement recommenders, a search engine, an analytics stack using Spark, and video playback plugins.

## 3.2 Methodological Challenges of Microservices

A major challenge with microservices is that one cannot simply rely on the client to report performance, as with traditional client-server applications. Resolving performance issues requires determining which microservice is the culprit of a QoS violation, which typically happens through distributed tracing. We developed a distributed tracing system that records latencies at RPC granularity using the Thrift timing interface. RPCs and REST requests are timestamped upon arrival to and departure from each microservice, and data is accumulated, and stored in a centralized Cassandra database. We additionally track the time spent processing network requests, as opposed to application computation. In all cases the overhead from tracing is negligible, less than 0.1% on $99^{th}$ percentile latency, and 0.2% on throughput [20].

# 4 Evaluation

## 4.1 Scalability and Query Diversity

Fig. 3 shows the throughput-tail latency ($99^{th}$ percentile) curves for representative operations of the *Movie Streaming* service, when running on our local server cluster of two-socket 40-core Intel Xeon servers (E5-2660 v3), each with 128GB memory, connected to a 10GBps ToR switch with 10Gbe NICs. All servers are running Ubuntu 16.04, and power management and turbo boosting are turned off. To avoid the effects of interference between co-scheduled applications, we do not share servers between microservices for this experiment.

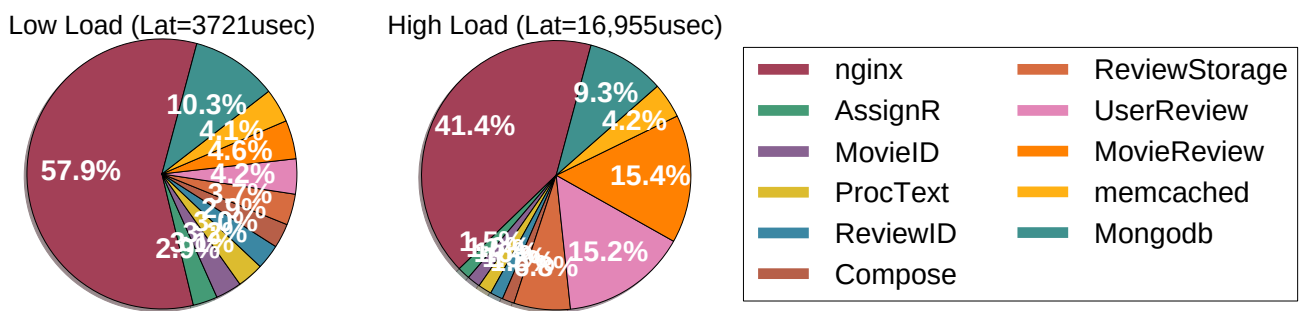

Fig. 4. Per-microservice breakdown for the *Movie Streaming* service.

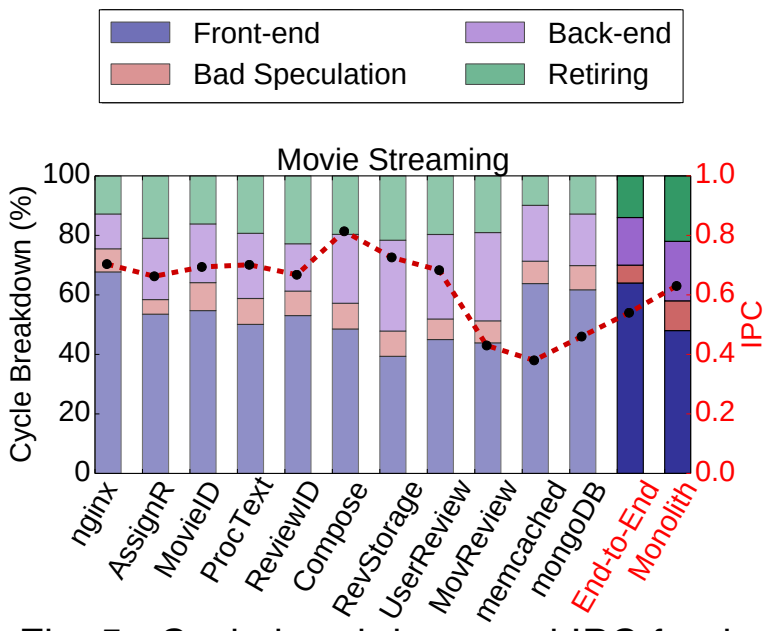

Fig. 5. Cycle breakdown and IPC for the *Movie Streaming* service.

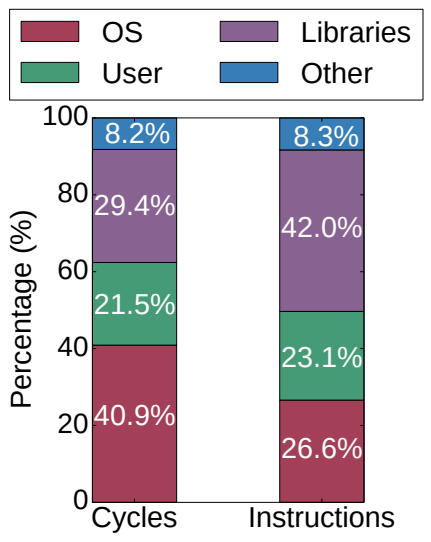

Fig. 6. Cycle and instructions breakdown to kernel, user, and libraries.

All experiments are repeated 10 times and the whiskers correspond to the $10^{th}$ and $90^{th}$ percentiles across runs.

We examine queries that browse the site for information on movies, add new movie reviews, and rent and stream a selected movie. Across all three request types the system saturates following queueing principles, although requests that process payments for renting a movie incur much higher latencies, and saturate at much lower load compared to other requests, due to the high bandwidth demands of streaming large video files. The latency curve for queries that browse movie information is also somewhat erratic, due to the variance in the amount of data stored across movies. The dataset consists of the 1,000 top-rated movie records, mined from IMDB, ca. 2018-01-31.

### 4.2 Implications in Server Design

**Cycles breakdown per microservice:** Fig. 4 shows the breakdown of the end-to-end latency across microservices at low and high load for *Movie Streaming*. We obtain the per-microservice latency using our distributed tracing framework, and confirm the execution time for each microservice with Intel's vTune. At each load level we provision microservices to saturate at similar loads, in order to avoid a single tier bottlenecking the end-to-end service, and causing the remaining microservices to be underutilized. At the moment this allocation space exploration happens offline, however we are exploring practical approaches that can operate in an online and scalable manner as part of current work. At low load, the front-end (`nginx`) dominates latency, while the rest of microservices are almost evenly distributed. MongoDB is the only exception, accounting for 10.3% of query execution time. This picture changes at high load. While the front-end still contributes considerably to latency, overall performance is now also limited by the back-end databases and the microservices that manage them (*ReviewStorage*, *MovieReview*, *UserReview*, *MongoDB*). This shows that bottlenecks shift across microservices as load increases, hence resource management must be agile, dynamic, and able to leverage tracing information to track how per-microservice latency changes over time. Given this variability across microservices we now examine where cycles are spent within individual microservices and across the end-to-end application.

**Cycles breakdown and IPC:** Fig. 5 shows the IPC and cycles breakdown for each microservice using Intel vTune [2], factoring in any multiplexing in performance counter registers. A large fraction of cycles, often the majority, is spent in the processor front-end. Front-end stalls occur for several reasons, including backpressure from long memory accesses. This is consistent with studies on traditional cloud applications [12], [16], although to a lesser extent for microservices than for monolithic services (*memcached*, *MongoDB*), given their smaller code footprint. The majority of front-end stalls are due to data *fetches*, while branch mispredictions account for a smaller fraction of stalls for microservices than for other interactive applications, either cloud or IoT [12], [23]. Only a small fraction of total cycles goes towards committing instructions, 22% on average, denoting that current systems are poorly provisioned for microservices-based applications. The same end-to-end service built as a monolithic Java application providing the exact same functionality, and running on a single node experiences significantly reduced front-end stalls, due to the lack of network requests, which translate to an improved IPC. Interestingly the cycles that go towards misspeculation are increased in the monolith compared to microservices, potentially due to its larger, more complex design, which makes speculation less effective. We see a similar trend later when looking at i-cache pressure (Fig. 10).

This analysis shows that each microservice experiences different bottlenecks, which makes generally-applicable optimizations, e.g., acceleration, challenging. The sheer number of different microservices is also a roadblock for creating custom accelerators. In order to find acceleration candidates we examine whether there is common functionality across microservices, starting from the fraction of execution time that happens at kernel versus user mode.

**OS vs. user-level cycles breakdown:** Fig. 6 shows the breakdown of cycles and instructions to *kernel*, *user*, and *libraries* for *Movie Streaming*. A large fraction of execution happens at kernel mode, and an almost equal fraction goes towards libraries like *libc*, *libgcc*, *libstdc*, and *libpthread*. The high number of cycles spent at kernel mode is not surprising, given that applications like memcached and MongoDB spend most of their execution time in the kernel to handle interrupts, process TCP packets, and activate and schedule idling interactive services [19]. The high number of library cycles is also justified given that microservices optimize for speed of development, and hence leverage a lot of existing libraries, as opposed to reimplementing the functionality from scratch.

**Computation:communication ratio:** After the OS, the network stack is a typical bottleneck of cloud applications [3]. Fig. 9 shows the time spent in network processing compared to application computation at low and high load for each microservice in *Movie Streaming*. At low load, TCP corresponds to 5-70% of execution time. This is a result of many microservices being too short to involve considerable processing, even at low load. At high load, the time spent queued and processing network requests dominates, with TCP processing bottlenecking the microservices responsible for managing the back-end databases. Microservices additionally shift the computation to communication ratio in cloud applications significantly compared to monoliths. For example, the same application built as a Java/JS monolith spends 18% of time processing client network requests, as opposed to application processing at low load, and 41% at high load. Despite the increased pressure in the network fabric, microservices allow individual components to scale independently, unlike monoliths, improving elasticity, modularity, and abstraction. This can be seen by the higher tail latency of the monolith at high load, despite the multi-tier application's complex dependencies.

**Brawny vs. wimpy cores:** There has been a lot of work on whether small servers can replace high-end platforms in the cloud [14], [15]. Despite the power benefits of simple cores, however, interactive services still achieve better latency in machines that optimize for single-thread performance. Microservices offer an appealing target for small cores, given the limited amount of computation per microservice. We evaluate low-power machines in two ways. First, we use RAPL on our local cluster to reduce the frequency at which all microservices run. Fig. 7 shows the change in tail latency as load increases, and as the operating frequency decreases for the end-to-end service. We compare these results against four traditional monolithic cloud applications (`nginx`, `memcached`, `MongoDB`, `Xapian`). As expected, most interactive services are sensitive to frequency scaling. Among the monolithic workloads, MongoDB is the only one that can tolerate almost minimum frequency at maximum load, due to it being I/O-bound. The other three monoliths start experiencing increased latency as frequency drops, `Xapian` being the most sensitive [17], followed by `nginx`, and `memcached`. Looking at the same study for *Movie Streaming* reveals that, despite the higher tail latency of the end-to-end service, microservices are much more sensitive to poor single-thread performance than traditional cloud applications. Although initially counterintuitive, this result is not surprising, given the fact that each individual microservice must meet much stricter tail latency constraints compared to an end-to-end monolith, putting more pressure on performance predictability.

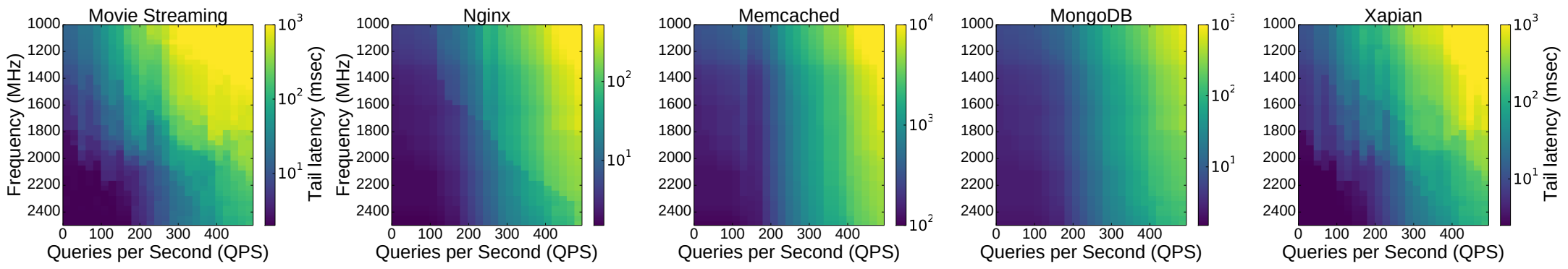

Fig. 7. Tail latency with increasing input load (QPS) and decreasing frequency (using RAPL) for the end-to-end *Movie Streaming* service, and for four traditional, monolithic cloud applications. The impact of reduced frequency is significantly more severe for *Movie Streaming*, as increased latency compounds across microservices.

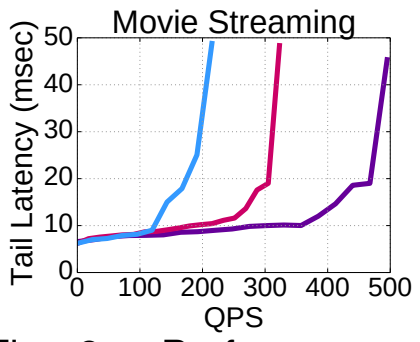

Fig. 8. Performance on a Xeon server at 2.6GHz (purple) and 1.8GHz (red), and on ThunderX (blue).

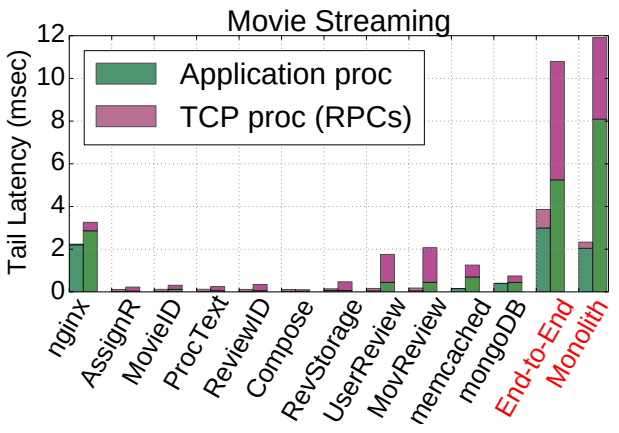

Fig. 9. TCP versus application processing for *Movie Streaming*.

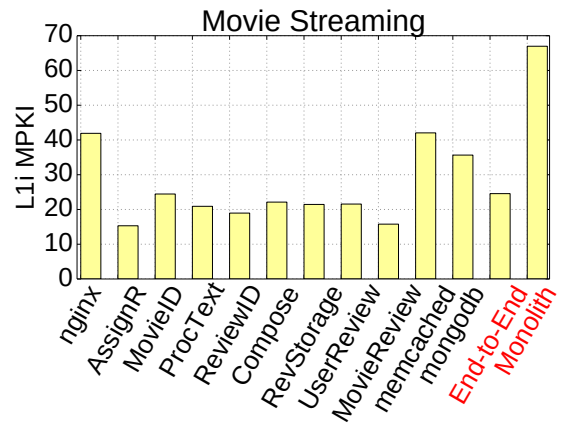

Fig. 10. Per-microservice L1-i MPKI for *Movie Streaming*.

Apart from frequency scaling, there are also platforms designed with low-power cores to begin with. We evaluate the end-to-end service on two Cavium ThunderX boards (2 sockets, 48 1.8GHz in-order cores per socket, and a 16-way shared 16M LLC). The boards are connected on the same ToR switch as the rest of our cluster, and their memory, network, and OS configurations are the same as the other servers [5]. Fig. 8 shows the throughput-latency curves for the two platforms. Although ThunderX is able to meet the end-to-end QoS at low load, it saturates much earlier than the high-end servers. We also show performance for Xeon at 1.8GHz which, although worse than the nominal frequency, still outperforms ThunderX by a considerable amount. Low-power machines can still be used for microservices out of the critical path, or insensitive to frequency scaling by leveraging the per-microservice characterization above.

**I-cache pressure:** Prior work has quantified the high pressure cloud services put on the instruction cache [12], [18]. Since microservices decompose what would have been one large binary to many loosely-connected services, we examine whether these results still hold. Fig. 10 shows the MPKI of each microservice in *Movie Streaming*. We also include the backend caching and database layers for comparison. First, the i-cache pressure of nginx, memcached, and MongoDB remains high, consistent with prior work. The i-cache pressure of the remaining microservices is considerably lower, which is expected given the microservices' small code footprints. `node.js` applications outside the context of microservices do not have low i-cache miss rates [23], hence we conclude that it is the simplicity of microservices which results in better i-cache locality. Most i-cache misses in *Movie Streaming* happen in the kernel, and using vTune, are attributed to the Thrift framework. In comparison, the monolithic design experiences extremely high i-cache misses, due to its large code footprint, and consistent with prior studies of cloud applications [18]. We also examined the LLC and D-TLB misses, and found them considerably lower than for traditional cloud applications, which is consistent with the push for microservices to be mostly stateless.

## 5 Conclusions

In this paper we highlighted the implications microservices have on system bottlenecks and datacenter server design. We used a new end-to-end movie reviewing and streaming service built with tens of microservices to quantify the instruction-cache pressure microservices create, the trade-off between big and small servers, and the shift they introduce in the ratio between computation and communication. As more cloud and IoT applications switch to this new application model, it is increasingly important to revisit the assumptions cloud systems have been previously built and managed with.